\documentclass[5p,times,twocolumn]{elsarticle}

\usepackage{amssymb}
\usepackage{latexsym}

\usepackage{amsmath}
\usepackage{lineno,hyperref,multirow}
\usepackage{booktabs}
\usepackage{makecell}

\journal{arXiv}

\begin{document}


\begin{frontmatter}

\title{Spatiotemporal implicit neural representation for unsupervised dynamic MRI reconstruction}
            
\author[a]{Jie Feng}
\author[a]{Ruimin Feng}
\author[b]{Qing Wu}
\author[a]{Zhiyong Zhang}
\author[b,c]{Yuyao Zhang}
\author[a]{Hongjiang Wei \corref{cor1}}

\address[a]{School of Biomedical Engineering, Shanghai Jiao Tong University, Shanghai, China}
\address[b]{School of Information Science and Technology, ShanghaiTech University, Shanghai, China}
\address[c]{iHuman Institute, Shanghaitech University, Shanghai, China}

\cortext[cor1]{Corresponding author.}
\ead{hongjiang.wei@sjtu.edu.cn}

\date{}

\begin{abstract}
Supervised Deep-Learning (DL)-based reconstruction algorithms have shown state-of-the-art results for highly-undersampled dynamic Magnetic Resonance Imaging (MRI) reconstruction. However, the requirement of excessive high-quality ground-truth data hinders their applications due to the generalization problem. Recently, Implicit Neural Representation (INR) has appeared as a powerful DL-based tool for solving the inverse problem by characterizing the attributes of a signal as a continuous function of corresponding coordinates in an unsupervised manner. In this work, we proposed an INR-based method to improve dynamic MRI reconstruction from highly undersampled \textbf{\textit{k}}-space data, which only takes spatiotemporal coordinates as inputs. Specifically, the proposed INR represents the dynamic MRI images as an implicit function and encodes them into neural networks. The weights of the network are learned from sparsely-acquired (\textbf{\textit{k}}, t)-space data itself only, without external training datasets or prior images. Benefiting from the strong implicit continuity regularization of INR together with explicit regularization for low-rankness and sparsity, our proposed method outperforms the compared scan-specific methods at various acceleration factors. E.g., experiments on retrospective cardiac cine datasets show an improvement of 5.5 $ \sim $ 7.1 dB in PSNR for extremely high accelerations (up to 41.6 $\times$). The high-quality and inner continuity of the images provided by INR has great potential to further improve the spatiotemporal resolution of dynamic MRI, without the need of any training data.
\end{abstract}


\end{frontmatter}


\section{Introduction}

Dynamic Magnetic Resonance Imaging (MRI) is one of the most popular MRI technologies, which can preserve not only excellent tissue contrast but also dynamic temporal changes of tissue. Dynamic MRI requires rapid data collection for the study of moving organs with severe physiological motion, such as the heart \citep{marcu2006clinical} and abdomen \citep{LOW2007525}. Dynamic Contrast-Enhanced (DCE) MRI has also made tremendous contributions to the study of microvascular structure and function of in vivo organs \citep{CUENOD20131187}.

However, the limitations of MRI hardware on gradient encoding performance and long acquisition time slow down our pace for higher spatiotemporal resolutions in dynamic MRI\citep{nbm1399}. Spatial and temporal resolution are always inversely related. High spatial resolution images can only be acquired with low temporal resolution and vice versa. Thus, a trade-off has to be made between spatial and temporal resolution in practical dynamic MRI. This conflict can be potentially resolved by developing advanced MRI reconstruction methods from highly-undersampled \textbf{\textit{k}}-space data, including the traditional Compressed-Sensing (CS)-based methods and the Deep-Learning (DL)-based methods. 

CS methods exploit spatial and temporal correlations of dynamic MRI by using irregular \textbf{\textit{k}}-space undersampling patterns to create incoherent artifacts in a suitable transform domain where the medical images are compressible, such as in the \textbf{\textit{k}}-t domain \citep{mrm21757},  temporal-gradient domain (temporal total variation regularizer) \citep{mrm24440, MRM24980} and many others. Image reconstruction is performed by exploiting the sparsity in the solution, subject to data consistency constraints. The further development of sparsity extended to the usage of low-rank prior: the Low-rank and Sparsity (L$\&$S) strategy enforced a both sparse and low-rank output solution \citep{5705578, 6214613}, and the Low-rank plus Sparsity (L+S) strategy decomposed the solution images into a low-rank and a sparsity component for background and the dynamic foreground, respectively \citep{mrm25240}. Recently, the subspace-modeling strategy enforced a combination of a temporal sparsity constraint and a low-rank spatial subspace constraint to improve DCE-MRI reconstruction \citep{feng2020grasp,feng20224d}.

Recent advances in DL techniques have shown potential for further accelerating dynamic MRI data acquisition. By adopting the supervised-learning strategy with large quantities of undersampled and fully-sampled image pairs, DL-based methods showed superior performance compared to CS-based methods \citep{bustin2020compressed}. DL-based methods applied in dynamic MRI reconstruction can be separated into two categories, i.e., end-to-end and unrolled methods. The end-to-end methods \citep{wang2016accelerating, mrm27106} enable the networks to directly learn the mapping from undersampled images with artifacts to fully sampled high-quality images. In contrast, the unrolled strategy is inspired by unrolling the iterative optimization process of CS, using networks to learn the auxiliary parameters or regularizers \citep{schlemper2017deep, qin2018convolutional, mrm28420, huang2021deep} during the iterations. Especially, L+S-Net \citep{huang2021deep} combined the L+S strategy of CS-based methods with the unrolled DL methods, demonstrating the availability of low-rank and sparsity in DL methods. However, the excessive demand for high-quality ground-truth labels in supervised learning hinders its applications in practice due to the generalization issue \citep{bustin2020compressed}. For example, the performance of the trained networks would degrade when the data is acquired with different scan parameters or pathological conditions. While in the case of DCE MRI, the ground-truth data are not available \citep{dce_sim}. Alternatively, the unsupervised-learning strategy was introduced to the DL-based dynamic MRI reconstruction without involving external data in the training process. For example, \citet{ke2020unsupervised} used a time-interleaved acquisition scheme, where the fully-sampled images were generated by merging adjacent frames. However, a large dataset is still needed for training the neural net. \citet{9442767} and \citet{9759446} both adopted the Deep Image Prior (DIP) approach \citep{Ulyanov_2018_CVPR}, which leveraged the tendency of untrained Convolutional Neural Networks (CNN) to generate natural-structured images as an implicit regularizer and then optimized the CNN parameters for scan-specific reconstruction. However, DIP-based methods suffer from a heavy computational burden and are still limited for application \citep{9442767}.  

Implicit Neural Representation (INR) is a new way which parameterizes signals by a multi-layer perceptron (MLP) \citep{SIREN}. Unlike traditional explicit representation that uses discrete elements such as pixels (2D images) or voxels (3D volumes), INR represents the desired object itself as a continuous representation function of the spatial coordinates. In other words, the values at any spatial location of the object can be retrieved by querying the trained MLP with the corresponding coordinate. It provides a general solution for various applications of object reconstruction. With the application of MLP and proper encoding function mapping the input coordinates to a high-dimensional space \citep{mildenhall2020nerf}, INR has achieved superior performance in multiple computer vision tasks \citep{park2019deepsdf,mildenhall2020nerf,muller2021real}. Previous research also showed the INR's capability to solve the inverse problem in medical imaging fields, e.g., CT image reconstruction 
\citep{zang2021intratomo,reed2021dynamic,9606601} and undersampled MRI  \citep{9788018} in an unsupervised manner. For example, implicit Neural Representation learning with Prior embedding (NeRP) \citep{9788018} was proposed to perform the static MRI reconstruction from the sparsely-sampled \textbf{\textit{k}}-space data. However, NeRP requires a fully-sampled prior image with the same modality for the reconstruction of longitudinal MRI images of follow-up scans. Additionally, the INR for object reconstruction usually takes hours or even days to converge on one single data. Recently, parametric encoding functions with extra learnable parameters \citep{liu2020neural,muller2021real,SunSC22,yu_and_fridovichkeil2021plenoxels} were proposed to significantly shorten the convergence time. For example, the hash encoding \citep{muller2021real} function has shown promising results for accelerating the computational processes of INR in seconds for many graphics applications. 

In this paper, we aim to present a new unsupervised method for highly accelerated dynamic MRI reconstruction. Inspired by the insight of INR, the proposed method treated the dynamic MR image sequence as a continuous function mapping the spatiotemporal coordinates to the corresponding image intensities. The function was parameterized by a hash encoding function and an MLP and served as an implicit continuity regularizer for dynamic MRI reconstruction. The MLP weights were directly learned from the imaging-model-based (\textbf{\textit{k}}, t)-space data consistency loss combined with the explicit regularizers, without training databases or any ground-truth data. When inferring, the reconstructed images can simply be querying the optimized  network with the same or denser spatiotemporal coordinates, which would allow for sampling and interpolating the dynamic MRI at an arbitrary frame rate. Experiments on retrospective cardiac cine data and prospective untriggered DCE liver MRI data showed that the proposed method outperformed the compared scan-specific methods. Our results showed an improvement of 5.5 dB $ \sim $ 7.1 dB in PSNR at an extremely high acceleration factor (41.6-fold). A temporal super-resolution test (4$\times$) was conducted without retraining the network to demonstrate the strong continuity of the optimized representation function as an implicit regularizer for dynamic MRI reconstruction. The main contributions of this study are as follows: 
\begin{itemize}
  \item INR is first introduced to dynamic MRI reconstruction as an implicit continuity regularizer, achieving an improvement of 5.5 dB $ \sim $ 7.1 dB in PSNR at an extremely high acceleration rate (41.6 $\times$) compared to other scan-specific methods. 
  \item The INR-based method is an unsupervised-learning strategy, meaning that it does not require external datasets or prior images for training. Thus, the proposed method generalizes on the data acquired with different scan parameters and imaging areas.
  \item 	The proposed method achieved a reasonable 4$\times$ temporal super-resolution for dynamic MRI reconstruction without network retraining, suggesting its strong implicit continuity to achieve higher temporal resolutions.
\end{itemize}

\section{Method}

\subsection{Dynamic MRI with regularizers}
In dynamic MRI, the relationship between measured (\textbf{\textit{k}}, t)-space data and the reconstructed image matrix can be expressed by a linear model. Given the discretized image matrix $d \in \mathbb{C}^{(N \times N) \times T}$ and the measured (\textbf{\textit{k}}, t)-space data of the $c$th coil $m_c \in \mathbb{C}^{(N \times M) \times T}$ ($1 \leq c \leq C$), where $N$ is the image size,$T$ denotes the total temporal frames of the image, $M$ ($M < N$) is the number of acquired readout lines for each frame and $C$ is the total number of coil channels. The relationship between $d$ and $m_c$ can be formulated as:
\begin{equation}\label{forward}
  m_c = F_uS_cd.
\end{equation}
Here, $F_u \in \mathbb{C}^{(N \times M) \times (N \times N)}$ denotes the Fourier operator with the undersampling mask, which simulates the undersampled acquisition process of dynamic MRI, and $S_c \in \mathbb{C}^{(N \times N) \times (N \times N)}$ is a diagonal matrix representing the $c$th coil sensitivity map.

Reconstructing image $d$ from the undersampled (\textbf{\textit{k}}, t)-space data is actually solving an ill-posed inverse problem, and the optimization process is formulated as:
\begin{equation}\label{eq2}
  \mathop{\arg\min}\limits_{d}{\frac{1}{2} \sum_{c = 1}^{C}  \Vert F_uS_cd-m_c \Vert_2^2 + \mathcal{R} (d)},
\end{equation}
where $\mathcal{R} (d)$ is the prior regularizer, helping target $d$ reach optimal results at ill-posed conditions. 

It has been shown that using sparsity and low-rank regularizers as prior knowledge in CS-based \citep{5705578, 6214613, mrm25240} and DL-based methods \citep{huang2021deep} is able to reach state-of-the-art results for dynamic MRI reconstruction. An example can be formulated as:
\begin{equation}\label{formulation}
  \mathop{\arg\min}\limits_{d}{\frac{1}{2} \sum_{c = 1}^{C}  \Vert F_uS_cd-m_c \Vert_2^2 + \lambda_S \Vert TV_t(d) \Vert_1 + \lambda_L \Vert d \Vert_*} ,
\end{equation}
where $TV_t(\bullet)$ is the temporal TV operator as the sparsity regularizer. $\Vert d \Vert_*$ is the nuclear norm (sum of singular values) of image matrix $d$, representing the low-rank regularizer. $\lambda_S$ and $\lambda_L$ are the sparsity and low-rank regularization hyperparameters, respectively. Previous works \citep{5705578} have proved that the target in Eq. \ref{formulation} can be optimized iteratively for a good dynamic MRI performance without Ground Truth (GT).

\subsection{INR in dynamic MRI}\label{theory}

\begin{figure*}[t]
  \centering
  \includegraphics[scale=1]{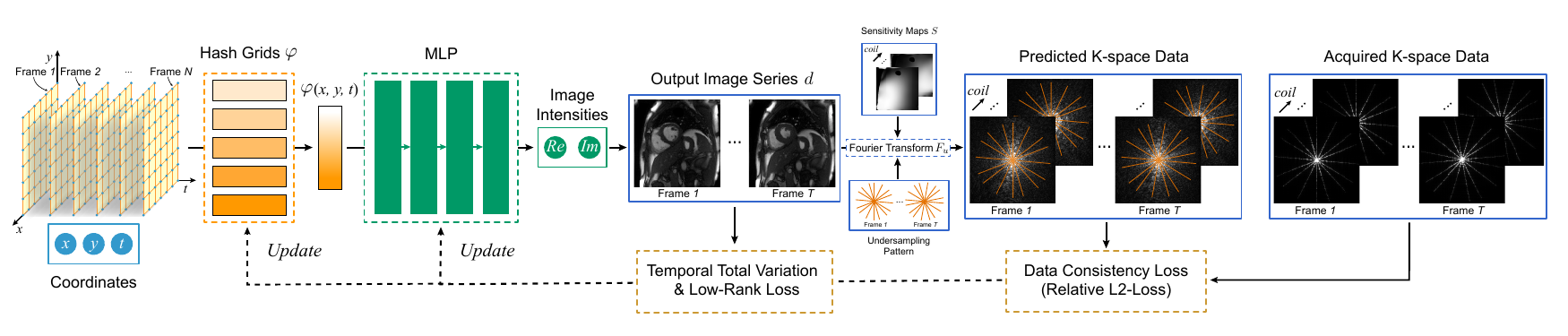}
  \caption{
    Overview of the proposed method. All the spatiotemporal coordinates are fed into hash grids and an MLP to output two-channel intensities as the real and imaginary parts of the image series. The predicted \textbf{\textit{k}}-space data are generated with the undersampled Fourier Transform (a golden-angle radial undersampling pattern) from the reconstructed complex-valued images following Eq. \ref{forward}. The difference between the predicted \textbf{\textit{k}}-space data and acquired \textbf{\textit{k}}-space data is calculated as the data consistency loss. Two regularization terms, temporal Total Variation and low-rankness, are applied to the output image series in the loss function. The parameters in the hash grids and the MLP are updated iteratively by minimizing the loss function. 
    }
  \label{fig1}
\end{figure*}

Inspired by INR, the internal continuity of the image can be a powerful regularizer for solving the ill-posed inverse problem of dynamic MRI reconstruction from sparsely-acquired (\textbf{\textit{k}}, t)-space data. The INR-based method can be implemented by applying a learnable continuous mapping function between spatiotemporal coordinates and desired image intensities to be reconstructed. We introduce $f_\theta: \mathbb{R}^3 \rightarrow \mathbb{C}$ be the continuous function parameterized by learnable parameters $\theta$, mapping the spatiotemporal coordinates $(x, y, t)$ into corresponding image intensities, where $(x, y)$ represent the 2D spatial coordinates $(1 \leq x, y \leq N)$ and $t$ represents the temporal coordinate $(1 \leq t \leq T)$. Thus, the image $d$ is rewritten to $d_\theta \in \mathbb{C}^{(N \times N) \times T}$ by feeding all the spatiotemporal coordinates of the dynamic images into $f_\theta$ and the Casorati matrix $d_\theta$ is:
\begin{equation}
  d_\theta = \begin{bmatrix}
    f_\theta(1, 1, 1) & \cdots & f_\theta(1, 1, T) \\
    \vdots &  & \vdots \\
    f_\theta(N, 1, 1) & \ddots & f_\theta(N, 1, T) \\
    \vdots &  & \vdots \\
    f_\theta(N, N, 1) & \cdots & f_\theta(N, N, T) 
  \end{bmatrix}.
\end{equation}

Thus, Eq. \ref{formulation} can be written as a fitting problem that searches the optimal parameters $\theta$ of the continuous mapping function $f_\theta$:
\begin{equation}\label{true_form}
  \mathop{\arg\min}\limits_{_\theta}{\frac{1}{2} \sum_{c = 1}^{C}  \Vert F_uS_cd_\theta-m_c \Vert_2^2 + \lambda_S \Vert TV_t(d_\theta) \Vert_1 + \lambda_L \Vert d_\theta \Vert_*}.
\end{equation}
Here, Eq. \ref{true_form} incorporates the implicit continuity on the desired image sequence, together with the explicit sparsity and low-rankness regularizers. 

\subsection{Continuous mapping function with MLP and hash encoding} \label{INR function}

In INR, the continuous representation function $f_\theta$ is based on MLP. A better high-frequency fitting performance can be achieved by mapping the input coordinates to a higher dimensional space using an encoding function $\varphi$ before passing them to MLP \citep{mildenhall2020nerf, tancik2020fourfeat}: 
\begin{equation}
  f_\theta(x, y, t)=MLP(\varphi(x, y, t)).
\end{equation}

In this work, we adopted hash encoding \citep{mueller2022instant} as the coordinate encoding function $\varphi$, which enables the use of smaller MLPs and significantly a faster convergence time. Specifically, hash encoding uses a total of $L$ independent hash grids with the size of $T$ as learnable feature storages. These hash grids represent a set of resolutions in the form of a geometric series, i.e., $N_{min}, b*N_{min}, \cdots, b^{(L-1)}*N_{min}$, where $N_{min}$ and b are the first term and the ratio of the geometric series, respectively. Trilinear interpolation is applied in each queried hash grid entry to keep continuity. Each hash grid outputs an $F$-dim feature vector and then these interpolated feature vectors are concatenated as the final encoded input vector. As pointed out by \citet{mueller2022instant}, the five hyperparameters mentioned above can be tuned to fit large quantities of tasks better: $N_{min}$ and $b$ decide how the resolution among different hash grids increases, and $L$, $T$, $F$ are important tuners for the tradeoff between performance, memory and quality. 

\subsection{Loss functions}
Eq. \ref{true_form} is rewritten to the form of the following loss functions for the implementation with gradient-descent-based algorithms:
\begin{equation}
  \mathcal{L}_{total} = \underbrace{\sum_{c = 1}^{C} \Vert F_uS_cd_\theta-m_c \Vert_2^2}_{\mathcal{L}_{DC}} + \lambda_S \underbrace{\Vert TV_t(d_\theta) \Vert_1}_{\mathcal{L}_{TV}} + \lambda_L \underbrace{\Vert d_\theta \Vert_*}_{\mathcal{L}_{LR}}.
\end{equation}
Here $\mathcal{L}_{DC}$, $\mathcal{L}_{TV}$ and $\mathcal{L}_{LR}$ stand for 
data consistency (DC) loss in (\textbf{\textit{k}}, t)-space, temporal TV loss and low-rank loss, corresponding to the three terms of the optimization objective in Eq. \ref{true_form}, respectively.

Considering that the magnitudes of the \textbf{\textit{k}}-space low-frequency elements are several orders greater than those of the high-frequency elements, a relative L2 loss \citep{lehtinen2018noise2noise, muller2021real} is used as the DC loss. Compared with normal L2 loss, the relative L2 loss is normalized by the square of predicted output, helping balance the gradients across \textbf{\textit{k}}-space for better high-frequency performance. Let $\hat{Y_i}$ be one element of the multi-coil predicted \textbf{\textit{k}}-space data 
$[FS_1d\ FS_2d\ \dots\ FS_Cd]$ and $Y_i$ is the corresponding element of 
the multi-coil acquired \textbf{\textit{k}}-space data $[m_1\ m_2\ \dots\ m_C]$, then DC loss is written as:
\begin{equation}
  \mathcal{L}_{DC} = \sum_{i=1}^{N \times M \times T \times C} \frac{(\hat{Y_i} - Y_i)^2}{(\hat{Y_i})^2 + \epsilon}.
\end{equation}
The parameter $\epsilon$ with a value of $10^{-4}$ is added to the denominator to prevent the zero-division problem.

Therefore, the parameters $\theta$ of hash grids and MLP are optimized to minimize the total loss:
\begin{equation}
  \mathcal{L}_{total} = \underbrace{\sum_{i=1}^{N \times M \times T \times C} \frac{(\hat{Y_i} - Y_i)^2}{(\hat{Y_i})^2 + \epsilon}}_{\mathcal{L}_{DC}} + \lambda_S \underbrace{\Vert TV_t(d_\theta) \Vert_1}_{\mathcal{L}_{TV}} + \lambda_L \underbrace{\Vert d_\theta \Vert_*}_{\mathcal{L}_{LR}}.
\end{equation}

\subsection{Implementation details}
We used a tiny MLP containing 5 hidden layers and each hidden layer consisted of 64 neurons followed by a ReLU activation function. The MLP output 2 channels, representing the real and imaginary components of the complex-valued MRI images. No activation function was adopted for the last layer. 

During the optimization process, all the spatiotemporal coordinates were gathered in one batch and the batch size was set to 1. All the coordinates were isotropically normalized to $\left[0,1\right]$ for fast convergence. The number of optimization epochs was set to 500. The Adam optimizer \citep{kingma2014adam} was used with a constant learning rate of 0.001, $\beta_1=0.9$, $\beta_2=0.999$, and $\epsilon={10}^{-8}$.

Once the optimization process was done, the continuous function $f_\theta$ was considered a good representation of the underlying image sequences. Then the same coordinate batch or a denser coordinate batch can be fed into the INR network to output the image sequences. 

The whole pipeline is illustrated in Fig.\ref*{fig1}, and was conducted on a system equipped with an Intel i7-9700 processor, 64G RAM, and an NVIDIA RTX 2080Ti 11G GPU. The networks were implemented with PyTorch 1.11.0 and tiny-cuda-nn \footnote{https://github.com/nvlabs/tiny-cuda-nn}. The non-cartesian Fourier undersampling operation was implemented with the Non-Uniform Fast Fourier Transform (NUFFT) and was deployed with torchkbnufft 1.3.0 \citep{muckley:20:tah} for fast calculation and gradient backpropagation on GPU. 

\section{Experiments and results}

\subsection{Setup}

\subsubsection{Datasets}
The proposed method was tested on a simulated retrospective cardiac cine dataset and a perspective untriggered DCE liver dataset to prove its effectiveness and generalization. 

(1)	Retrospective cardiac cine dataset:

The fully sampled cardiac cine data from the OCMR dataset \citep{chen2020ocmr} were acquired from healthy volunteers on a 1.5T scanner (MAGNETOM Avanto, Siemens Healthineers, Erlangen, Germany) using a bSSFP sequence with the following parameters: FOV = $320 \times 260$ mm$^2$, imaging matrix = $256 \times 208$, slice thickness = 8 mm, TR/TE = 2.79 ms/1.33 ms, number of frame = 18. The data acquisition was collected with prospective ECG-gating and breath-holding. The number of receiver coils is 18. A simulation undersampling pattern of 2D golden-angle radial acquisition scheme is adopted, where the readout lines are repetitively through the center of \textbf{\textit{k}}-space and rotated with a step of 111.25°. The simulation process includes cropping original data to $208 \times 208$ in the image domain and then converting to the frequency domain by multi-coil NUFFT with golden-angle trajectories of Fibonacci numbers \citep{chandarana2013free}. The coil sensitivity maps were calculated by the ESPIRiT algorithm \citep{ESPIRiT}.

(2) Untriggered DCE liver dataset:
  
The DCE liver data were acquired continuously with the golden-angle acquisition scheme. The 3D stack-of-stars Fast Low Angle SHot (FLASH) sequence was acquired on a breath-holding healthy volunteer using a 3T Siemens MAGNETOM Verio scanner with the following parameters: FOV = $370 \times 370$ mm$^2$, TR/TE = 3.83 ms/1.71 ms, imaging matrix = $384 \times 384$, slice thickness = 3 mm, total spoke number of each slice = 600. 
A total of 12 receiver coils were used during the scan. The data including coil sensitivity maps were from \citet{MRM24980}'s demo and details about intravenous contrast enhancement can be found in the paper. Each 34 acquired spokes were grouped to reconstruct one frame, which corresponds to 
an Acceleration Factor (AF) $\approx 11.3$ and 17 frames in total.

\subsubsection{Performance evaluation} \label{setup}
In this work, we chose NUFFT, L+S \citep{mrm25240} and GRASP \citep{MRM24980} as the baselines for comparison. NUFFT gives the results obtained by directly zero-filling the frequency domain. L+S and GRASP two of the CS-based reconstruction methods which use a similar optimization pipeline as Eq. \ref{eq2}. The difference between them is that GRASP adopted a temporal TV regularizer, while L+S decomposed the solution images into a background component with low-rank regularizer and a dynamic foreground with temporal TV regularizer. We did not compare the proposed INR-based method to the supervised DL methods for dynamic MRI reconstruction since the datasets used in this work are insufficient for supervised network training. In addition, the ground truth is not available for the untriggered DCE liver dataset, which also limits the training process of previous supervised methods.

We tested the performance of the proposed method with 21, 13, 8 and 5 spokes per frame (AF $\approx 9.9, 16, 26, 41.6$) on the cardiac cine dataset, and with 34 spokes per frame (AF $\approx 11.3$) on the DCE liver dataset. For a fair comparison, the hyperparameters of all the methods are tuned to get the best performance and fit the GPU storage in different datasets and AFs, respectively. 

Quantitative visual comparison and quantitative comparison were used for evaluation. For the cardiac cine dataset, quantitative metrics including peak signal-to-noise ratio (PSNR) and structural similarity index (SSIM) were calculated frame-by-frame as follows:
\begin{equation}
  \mathit{PSNR} = 10\times{log}_{10}{(\frac{1}{\left\lVert y-\hat{y} \right\rVert _2^2})},
\end{equation}
\begin{equation}
  \mathit{SSIM} =\ \frac{(2\mu_y\mu_{\hat{y}}+c_1)(2\sigma_{y\hat{y}}+c_2)}{({\mu_y}^2+{\mu_{\hat{y}}}^2+c_1)({\sigma_y}^2+{\sigma_{\hat{y}}}^2+c_2)}\ ,
\end{equation}
where $y$ and $\hat{y}$ represent ground truth and reconstructed image, respectively, $\mu_y$ and $\mu_{\hat{y}}$ are the mean intensity of $y$ and $\hat{y}$, $\sigma_y$ and $\sigma_{\hat{y}}$ are the variance of $y$ and $\hat{y}$, $\sigma_{y\hat{y}}$ is the covariance of $y$ and $\hat{y}$, the constant $c_1$ and $c_2$ were set to $0.01^2$ and $0.03^2$. $y$ and $\hat{y}$ were both normalized to $[0, 1]$ according to the image sequence maximum and minimum. 

The \textbf{\textit{k}}-space data were calculated from the reconstructed complex-valued MR images with the 2D Fast Fourier Transform. For quantitative comparison, the normalized root mean square error (NRMSE) against GT \textbf{\textit{k}}-space data was calculated coil-by-coil:
\begin{equation}
  \mathit{NRMSE} = \sqrt{\frac{\left\lVert Y-\hat{Y} \right\rVert _2^2}{\left\lVert Y \right\rVert _2^2} },
\end{equation}
where $Y$ and $\hat{Y}$ represents the predicted and acquired \textbf{\textit{k}}-space data, respectively. 

For the DCE liver dataset, only visual comparison and temporal ROI intensity assessment were conducted due to the lack of the GT image. The ROIs of the aorta (AO) and portal vein (PV) were manually drawn for signal intensity-time curves. For temporal fidelity comparison, NUFFT was used as the reference since no temporal regularization was involved in the reconstructed images. Although contaminated by the streaking artifacts, the average signal intensity from NUFFT results across large ROI was still able to preserve contrast evolution for fidelity analysis.

\subsection{Reconstruction performance of the proposed method}

\subsubsection{Cardiac cine dataset}
\begin{figure*}[!t]
  \centering
  \includegraphics[scale=0.75]{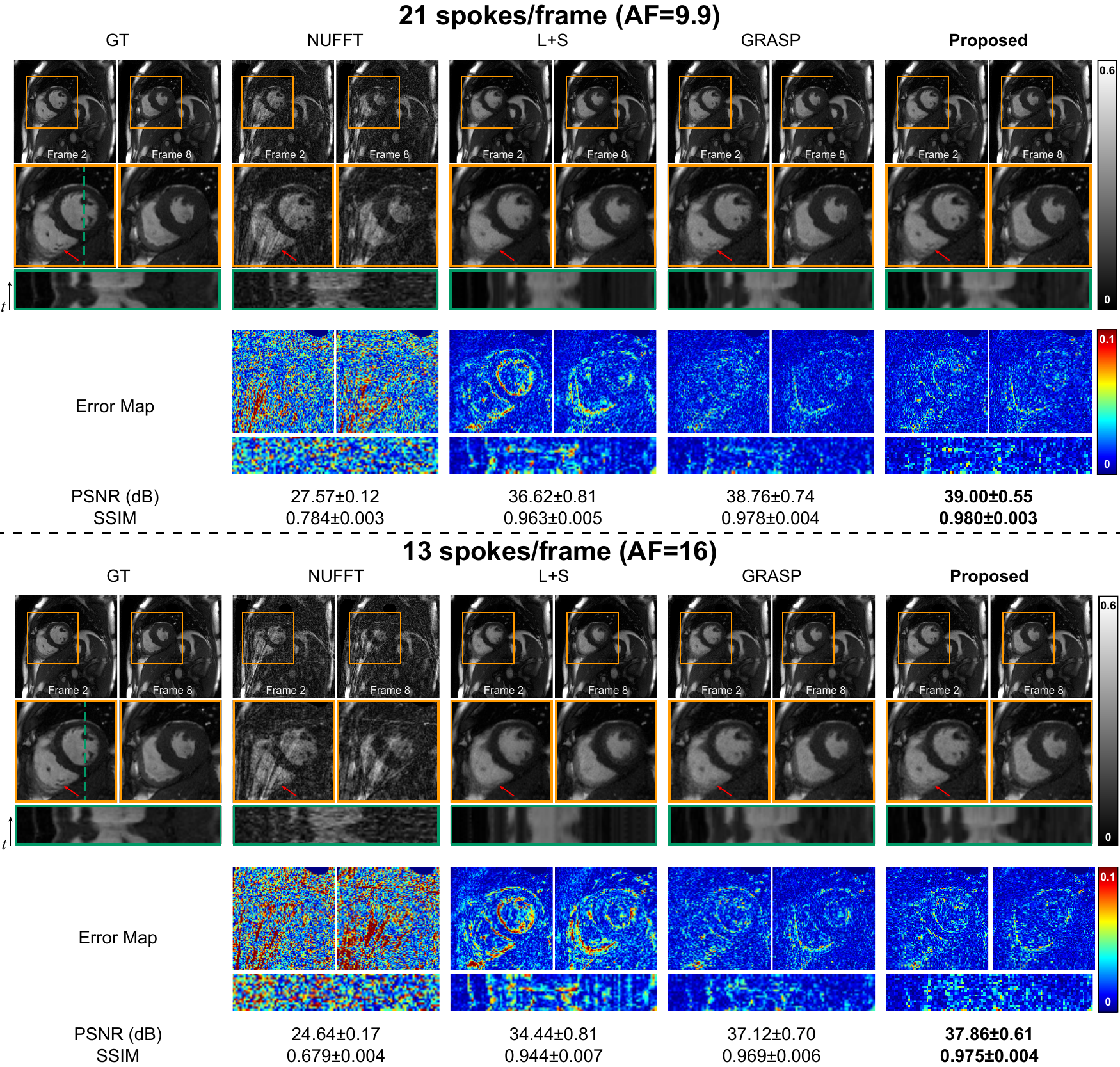}
  \caption{
    The reconstruction results of NUFFT, L+S, GRASP and the proposed method (from left to right) on the cardiac cine dataset with 21 and 13 spokes per frame (AF=9.9, 16). The enlarged views of the heart region are outlined by the orange boxes and the red arrows point out the structure where the proposed method gives a superior reconstruction performance. The y-t images (the 116th slice along y and temporal dimensions) are outlined by green boxes. The error maps and PSNR/SSIM metrics are shown at the bottom, respectively.
  }
  \label{fig2}
\end{figure*}

Fig.\ref{fig2} compares the reconstruction performance of different methods on the cardiac cine dataset with 21 and 13 spokes per frame (AF=9.9, 16, respectively). Visually, the images reconstructed by the proposed method appear to have better anatomical details and provide a more accurate temporal fidelity than the baselines at both acceleration conditions of 21 and 13 spokes. NUFFT, L+S and GRASP all suffer from artifacts and noises in the cardiac chamber area. While the reconstructed images by the proposed method show highly similar anatomical details as the ground truth, as pointed out by red arrows. In the y-t view, the L+S results are over smooth and the GRASP results suffer from noticeable streaking artifacts along the temporal axis. The proposed method provides the highest temporal fidelity of the dynamic images between frames. The error map between the reconstruction and the ground truth further supports our observation. It is noted that the reconstructed errors observed on the error maps at the edge of the cardiac ventricles were potentially blurred by cardiac motion. The proposed INR-based method shows the smallest error at the edge of the ventricles, and is consistent with the observation from the y-t view. Quantitatively, the proposed method achieves the best performance with a PSNR of 39.00±0.55 dB (21 spokes) / 37.86±0.61 dB (13 spokes) and an SSIM of 0.980±0.003 (21 spokes) / 0.975±0.004 (13 spokes) than the compared methods.

\begin{figure*}[!t]
  \centering
  \includegraphics[scale=0.75]{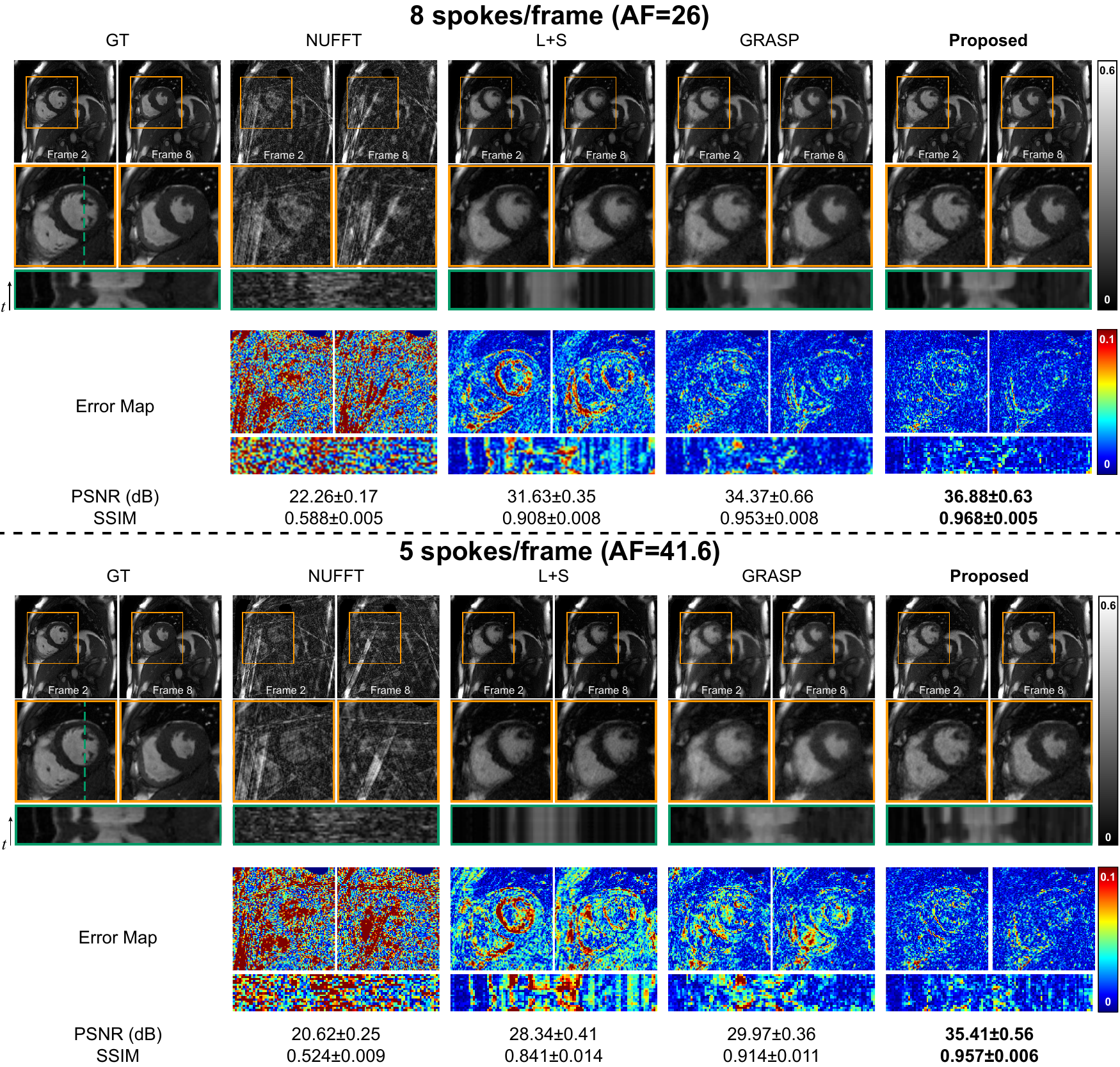}
  \caption{
    The comparison of the reconstruction results on the cardiac cine dataset with 8 and 5 spokes per frame, which corresponds to the acceleration factors of 26 and 41.6. Zoomed-in views of the heart chambers are outlined by orange boxes and the y-t images (the 116th slice along y and temporal dimensions) are outlined by green boxes. The difference map between the reconstructed image and ground truth and PSNR/SSIM metrics are also shown.
  }
  \label{fig4}
\end{figure*}
\begin{figure*}[!h]
  \centering
  \includegraphics[scale=0.85]{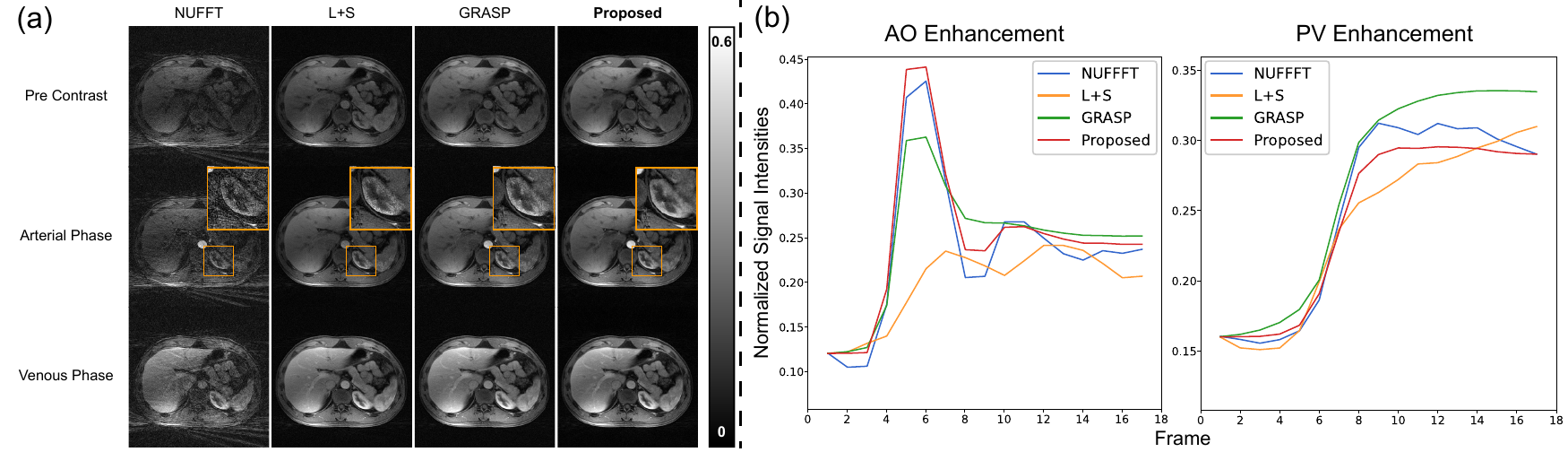}
  \caption{
    The comparison of the reconstruction results and ROI analysis among different methods on the DCE liver dataset with 34 spokes per frame (AF=11.3). (a) Reconstruction results at different contrast phases are visualized. The zoomed-in area outlined by orange boxes with the proposed method gives the best image quality with minimal noise among different methods. (b) Signal intensity-time curves of different methods are compared in aorta (AO) and portal vein (PV) areas, and the NUFFT result serves as the temporal fidelity reference.
    }
  \label{fig5}
\end{figure*}
We further tested the ability of the proposed method for dynamic MRI reconstruction at extremely high acceleration rates (8 and 5 per frame, AF=26 and 41.6, respectively), as shown in Fig.\ref{fig4}. The proposed method exhibits comparable performance between AF=26 and 41.6, with the PSNR/SSIM of 36.88 ± 0.63 dB/0.968 ± 0.005 (8 spokes) and 35.41 ± 0.56 dB/0.957 ± 0.006 (5 spokes). The proposed method has the best image quality with minimal noise and artifacts. Contrarily, L+S and GRASP suffer from temporal smoothness and noticeable streaking artifacts with increased acceleration rates. From the y-t view, the dynamic information on the reconstructed images is well captured by the proposed method, even with 5 spokes per frame. Additionally, the proposed INR-based method results in a higher PSNR than GRASP (~5.5 dB) and L+S (~7.1 dB), respectively.

\subsubsection{DCE liver dataset}
For DCE liver dataset with 34 spokes per frame (AF=11.3), the visual comparisons at different temporal phases are demonstrated in Fig. \ref{fig5}(a). As can be seen from the zoomed-in images, the anatomical details of the kidney can be well visible on the reconstructed images by the proposed method. Severe streaking and noise can be observed on the reconstructed images by NUFFT, L+S, and GRASP. While the proposed method provides high-quality images with less noise than other methods. The signal intensity-time curves in Fig. \ref{fig5}(b) suggest that the proposed method yields the best temporal fidelity, which is consistent with the results of NUFFT in AO and PV. For example, the intensity fluctuation of the AO curve between Frame 5 and Frame 11 can be well captured by the proposed INR-based method. 

\subsection{Results of the temporal super-resolution}
\begin{figure}[!h]
  \centering
  \includegraphics[scale=1]{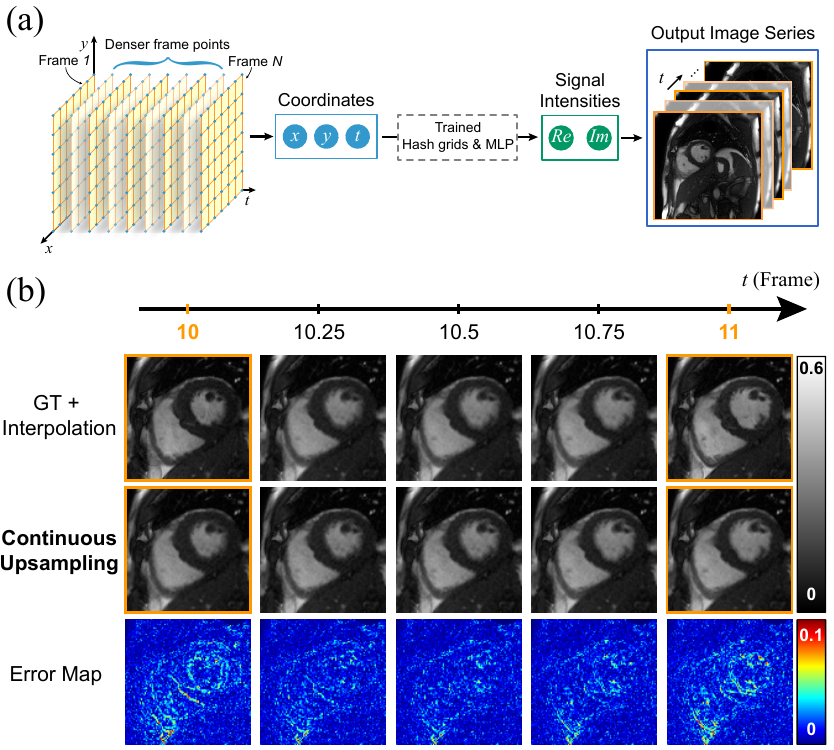}
  \caption{
    (a) The pipeline of temporal super-resolution for the reconstructed dynamic MRI. For the given denser coordinates, the optimized function (Hash grids $\&$ MLP) outputs the interpolated frames. (b) The upsampled images between Frame 10 and Frame 11 of the cardiac cine dataset with 21 spokes per frame. Three equally-spaced coordinates to be generated (10.25, 10.5, 10.75) between Frame 10 and Frame 11 are fed to the network for temporal super-resolution $(4\times)$. The ground truth of Frame 10 and Frame 11, and the linear interpolated frames serve as the reference. The reference and output images at the position of Frame 10 and 11 are outlined with orange boxes. The corresponding error maps are displayed at the bottom. 
    }
  \label{fig7}
\end{figure}
To demonstrate the internal continuity of the optimized representation of the dynamic MRI, we use a denser coordinate along the temporal axis as input to conduct upsampling $(4\times)$ on the reconstructed dynamic MR image sequence, named temporal super-resolution. The pipeline is shown in Fig.\ref{fig7}(a). The GT frames with temporal linear interpolation between Frame 10 and 11 are used as the reference for comparison, as shown in Fig.\ref{fig7}(b). Qualitatively, there is no significant structural difference between the super-resolution images and the interpolated images, indicating the strong implicit continuity representation of the optimized INR function. 

\section{Discussion}

In this study, we proposed a novel unsupervised INR-based deep learning method for highly accelerated dynamic MRI reconstruction, which modeled the dynamic MR image sequence as a continuous mapping function. We validated the proposed method on retrospective cardiac cine data and perspective DCE liver data with various acceleration rates. The results showed the effectiveness and generalization of the proposed method on artifact suppression and motion fidelity preservation, especially at extremely high accelerations of 26-fold or 41.6-fold. The proposed method outperforms the compared CS-based methods such as L+S and GRASP. The results indicated that the proposed reconstruction method holds promise for high temporal resolution 2D MRI acquisitions.

The superiority of the proposed method over the baseline methods is believed from the implicit regularization from the internal continuity of INR, which is validated by the results of the temporal super-resolution $(4\times)$, as shown in Fig.\ref{fig7}. In addition, the super-resolution performance allows us to further speed up the data acquisition along the temporal axis for dynamic MRI. Unlike the existing super-resolution methods, the INR-based method does not require extra modeling or training, but simply gives the denser coordinates, which reduces the computational burden and the reconstruction time usage during deployment. 

The proposed method has a few limitations. First, the low-rank regularization adopted in this work is the nuclear norm and is optimized with gradient-descent-based algorithms. However, as discussed by \citet{5705578}'s work, naive nuclear norm minimization may not be stable for fast convergence. In future works, INR combined with different low-rank regularization substitutes and optimization methods will be explored. Second, the temporal super-resolution test indicates a comparable 4 times upsampling result with INR, but the smoothness witnessed at the edge of heart chambers demonstrated its limitation for higher or even arbitrary super-resolution results. Third, although the reconstruction time is faster than the other unsupervised methods, it is still challenging for real-time imaging. 

\section{Conclusion}
In this work, we proposed an INR-based unsupervised deep learning method for highly accelerated dynamic MRI reconstruction. The proposed method learns an implicit continuous representation function to represent the desired spatiotemporal image sequence, mapping spatiotemporal coordinates to the corresponding image intensities. The proposed method is training database-free and does not require prior information for the reconstruction. Several tests on retrospective cardiac and perspective DCE liver data proved that the proposed method could robustly produce a high-quality dynamic MR image sequence even at an extremely high acceleration rate $(41.6 \times)$. Additionally, benefiting from the internal continuity of the optimized INR network, the proposed method demonstrates an impressive performance of temporal super-resolution to upsample the desired dynamic images at higher temporal rates than the physical acquisitions. We thus believe that the INR-based method has the potential to further accelerate dynamic MRI acquisition in the future. 

\bibliographystyle{model1-num-names}
\bibliography{refs}

\end{document}